\documentclass[aip,cha,reprint,twocolumn,nofootinbib,longbibliography,floatfix]{revtex4-2}

\usepackage{amsmath,amssymb,bm}
\usepackage{graphicx}
\usepackage{bm}
\usepackage{booktabs}
\usepackage{microtype}
\usepackage{xcolor}
\usepackage{enumitem}
\usepackage[colorlinks=true,citecolor=blue,urlcolor=blue,linkcolor=blue]{hyperref}

\hypersetup{
  colorlinks=true,
  linkcolor=blue!45!black,
  citecolor=blue!45!black,
  urlcolor=blue!55!black,
  pdftitle={NeuroAI Book-- Chapter 2: Neural Logic, Invariance, and the Retina---McCulloch and Pitts},
  pdfauthor={Nima Dehghani}
}

\graphicspath{{figures/}}
\newcommand{\dd}{\mathrm{d}}
\newcommand{\E}{\mathbb{E}}
\newcommand{\Prob}{\mathbb{P}}

\newcommand{\Cov}{\mathop{\mathrm{Cov}}}

\newcommand{\one}{\mathbb{I}}
\newcommand{\RR}{\mathbb{R}}

\begin{document}

\title[\emph{Dehghani, N. -- \textbf{NeuroAI}: Theoretical Foundation of Dynamics, Learning and Computation in Brains, Minds $\&$ Machines}]{Chapter 2: Neural Logic, Invariance, and the Retina---McCulloch and Pitts}
\author{Nima Dehghani}
\affiliation{McGovern Institute for Brain Research, Massachusetts Institute of
Technology, Cambridge, Massachusetts 02139, USA}
\affiliation{The NSF AI Institute for Artificial Intelligence and Fundamental
Interactions (IAIFI), Massachusetts Institute of Technology, Cambridge,
Massachusetts 02139, USA}
\email{nima.dehghani@mit.edu}
\date{2026}

\maketitle

\section*{Chapter orientation}

Within \textit{NeuroAI: Theoretical Foundations of Dynamics, Learning and
Computation in Brains, Minds, and Machines}\footnote{For the NeuroAI Book see:\\ \url{https://neurovium.science/books/NeuroAI}}, this chapter opens the book's
first part, on the formalization of neural computation, after the introductory
chapter has defined NeuroAI and separated it from the application of AI tools
to neural data. McCulloch and Pitts supply the first decisive formal move: a
material nervous system is treated as a causal network whose activity can
realize logical relations through threshold, inhibition, delay, and recurrence.
The next chapter replaces the fixed circuit by Rosenblatt's adaptive
perceptron. Amari then develops recurrent collective dynamics and the geometry
of learning, and Hopfield converts associative memory into motion on an energy
landscape. Together, these chapters form the foundational arc from neural
logic to learning, macroscopic dynamics, and many-body memory.

The chapter does not reduce that arc to the familiar cartoon of one binary
neuron. It reconstructs the 1943 distinction between excitatory summation and
inhibitory veto, Boolean synthesis, the physical meaning of logical delay, and
the finite-state consequences of recurrence. It then follows McCulloch and
Pitts beyond threshold logic: heterarchy turns directed cycles into an
obstruction to scalar value, the 1947 work on universals asks how a physical
network can quotient nuisance transformations, and the 1959 frog-retina study
uses single-fiber recordings to identify parallel, invariant operations at a
sensory output. The formal and experimental works are treated as parts of one
research program about how a physical knower preserves task-relevant
relations.

The mathematical route separates exact historical claims from later
formalization. A fixed recurrent network of $N$ binary units has at most
$2^N$ states; it is not thereby an unbounded Turing machine. A single weighted
threshold element computes only linearly separable Boolean functions, whereas
networks of such elements can synthesize every finite Boolean function. Group
averaging, equivariance, feedback canonicalization, and spike-triggered
analysis are used as modern tools without attributing their current notation
to the historical papers. Seven deterministic computational studies, generated
with fixed seed 1943, distinguish exact enumeration, qualitative effective
models, system-identification results, and original schematics.

\section{Why McCulloch belongs in a physics of neural computation}
\label{sec:why}

\begin{figure*}[t]
\centering
\includegraphics[width=.96\textwidth]{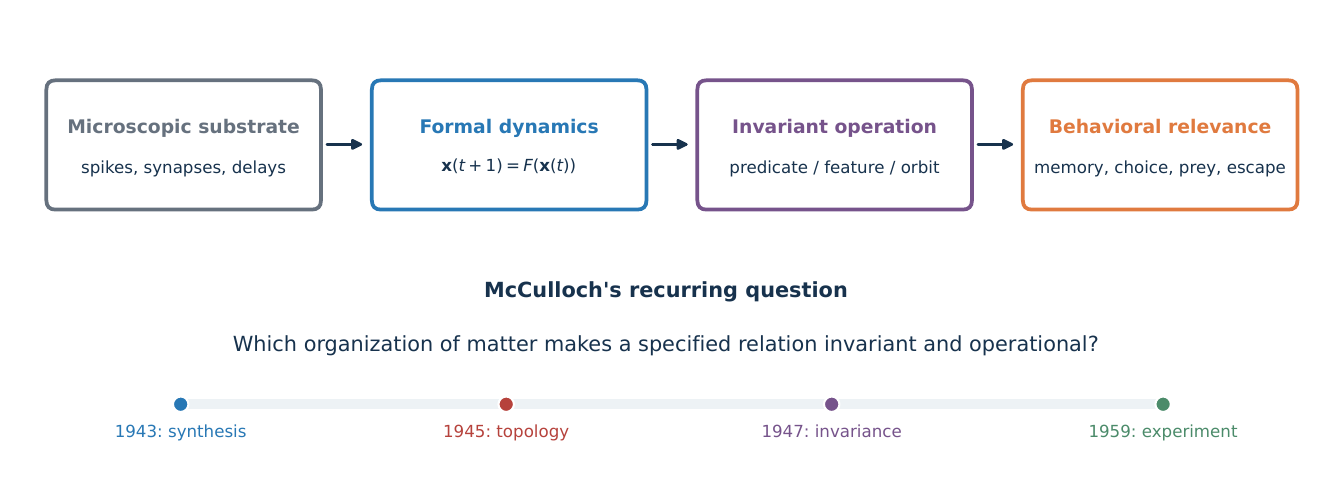}
\caption{The through-line of McCulloch's research. The 1943 paper synthesized
nets that realized specified propositions; the 1945 paper related circular
preference to network topology; the 1947 paper sought invariant operations; and
the 1959 frog study identified such operations experimentally. This is an
original schematic, not a claim that every paper followed a single linear plan.}
\label{fig:program}
\end{figure*}

McCulloch (1898--1969) was a neuropsychiatrist, neurophysiologist,
mathematical modeler, philosopher, and organizer of cybernetics. Those labels do
not merely describe breadth. They identify four scales at which he insisted a
theory of mind must close. There is a material substrate: neurons, synapses,
conduction, inhibition, and sensory anatomy. There is a dynamics that maps the
present physical state to a later one. There is an operation that remains
meaningful despite irrelevant physical variation. Finally, there is a relation
to memory, choice, recognition, or behavior. Figure~\ref{fig:program} shows this
recurring structure.

This belongs to a physics of neural computation because McCulloch treated cognition as organization
of matter in time. The important objects are state space, causal propagation,
feedback, invariants, and constraints imposed by topology. Unlike the later
Hopfield model, the 1943 network does not generally possess an energy function.
Unlike a Boltzmann machine, it has no temperature or equilibrium measure. Its
physics is instead that of deterministic switching systems and finite
dynamical machines. The frog paper changes the epistemic direction: rather
than prescribe a function and synthesize a net, it perturbs a living sensory
system and infers which functions its output fibers perform.

The historical models and modern descendants should be compared through their
operations, not through retrospective vocabulary. The 1943 unit is a discrete
causal element rather than a differentiable one; its propositional semantics
describes relations among neural events rather than a commitment to conscious
sentences. The frog study, in turn, established parallel feature-selective maps
without requiring the machinery of a modern convolutional network. Threshold
circuits, finite automata, invariance, canonicalization, and behaviorally chosen
stimuli are therefore real continuities, provided that the level of comparison
is kept explicit.

\subsection{Technical route and prerequisites}

The reader should know elementary graph theory, linear algebra, probability,
and ordinary differential equations. Sections~\ref{sec:logicalmodel}--
\ref{sec:power} form the 1943 logic and dynamics core. Work through the XOR
proof and the finite-state periodicity theorem before reading the discussion of
computational universality. Section~\ref{sec:heterarchy} isolates the topological
argument of 1945. Sections~\ref{sec:universals}--\ref{sec:canonical} develop the
1947 invariance program. Sections~\ref{sec:frog}--\ref{sec:identification}
move from the frog experiment to a modern effective model and system
identification.

All figures were generated by
\texttt{code/mcculloch\_replication.py}. They are controlled pedagogical
replications or original schematics, not digitizations of published figures.
The retinal model is deliberately labeled as a qualitative model; no raw 1959
recordings were fitted. The fixed random seed is 1943.

\section{Historical path: from psychons to experimental epistemology}
\label{sec:history}

McCulloch's early question was epistemological: what kind of event could serve
as an elementary proposition in a physical brain? The all-or-none action
potential suggested a bivalent event, while anatomy supplied directed
connections and recurrent paths. His training crossed philosophy, psychology,
medicine, psychiatry, and neurophysiology, and his temperament was equally
synthetic: he treated poetry, logic, anatomy, and engineering as routes into the
same question of how a material organism can know. At Chicago and later in
MIT's Building 20, he created unusually porous intellectual communities. That
organizational work was not incidental to cybernetics; it helped turn topics
that had belonged to separate disciplines into a shared scientific problem.

\begin{figure*}[t]
\centering
\includegraphics[width=.96\textwidth]{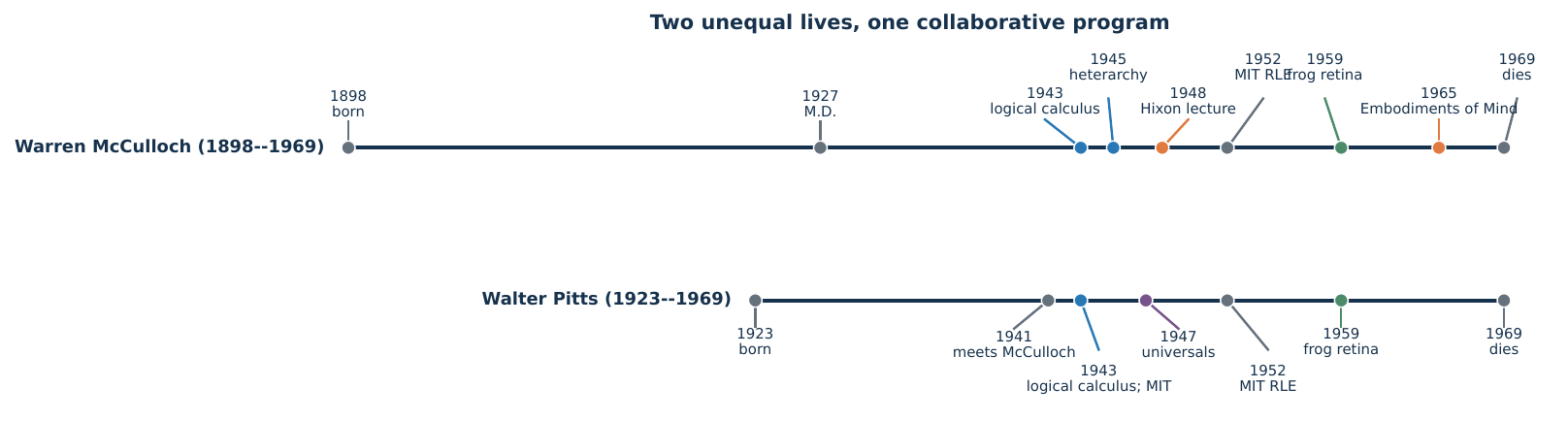}
\caption{Parallel selective timelines for Warren McCulloch and Walter Pitts.
Shared works appear on both tracks, while the different spans make visible how
much of Pitts's contribution was compressed into a short life. The Hixon marker
dates McCulloch's 1948 lecture; the Jeffress proceedings appeared in 1951.
Dates otherwise label births, deaths, publications, or institutional
transitions, not exclusive moments of invention. Original schematic.}
\label{fig:timeline}
\end{figure*}

Walter Pitts (1923--1969) supplied much of the collaboration's mathematical
force. A largely self-taught logician from Detroit, he met McCulloch through
Jerome Lettvin while still eighteen. McCulloch offered him a home and an
intellectual partnership; Pitts, in return, gave technical form to ideas about
recurrence, temporal reference, and logical realizability. He was not a junior
assistant hidden behind the compound name ``McCulloch--Pitts'': he was coauthor
of the 1943 calculus, first author of the 1947 universals paper, and a central
mathematical presence in the MIT group. His later withdrawal, destroyed
dissertation, and death at forty-six make the brevity of the lower timeline in
Figure~\ref{fig:timeline} painfully literal. Amanda Gefter's profile, ``The Man
Who Tried to Redeem the World with Logic,'' is an unusually vivid biographical
entry point; it is best read alongside Tara Abraham's archival history, which
provides a broader historical frame.\cite{Gefter2015,Abraham2016}

Their collaboration did not begin neural modeling from nothing. Rashevsky's mathematical biophysics,
Householder and Landahl's network models, neurophysiology, switching theory, and
mathematical logic were already active traditions.
Shannon had shown how Boolean algebra describes relay circuits, and Turing had
defined effective computation by an abstract machine.
\cite{Shannon1938,Turing1936,Abraham2002,Piccinini2004}

The distinctive 1943 synthesis joined a proposed neuron-level mechanism to a
logical calculus over time. The paper's two directions matter equally:

\begin{enumerate}[leftmargin=*]
\item given a net, describe the propositions realized by its activity; and
\item given an admissible logical expression, construct a net whose activity
realizes it.
\end{enumerate}

The first direction is analysis; the second is synthesis. Their combination
makes the paper more than a metaphor that a neuron is ``like'' a gate.
McCulloch's 1945 heterarchy paper then asked what recurrent topology implies for
choice when preferences are cyclic.
\cite{McCulloch1945} Pitts and McCulloch's 1947 paper asked how a nervous system
could recognize a form across transformations of pitch, size, and position.
\cite{PittsMcCulloch1947} McCulloch delivered his Hixon Symposium lecture in
September 1948; it was published in the Jeffress volume in 1951. Under the title
``Why the Mind Is in the Head,'' it joined information, reverberation, feedback,
prediction, and invariance.\cite{McCulloch1951}

McCulloch moved to MIT's Research Laboratory of Electronics in 1952. The 1959
frog paper, with Jerome Lettvin, Humberto Maturana, and Pitts, replaced a largely
synthetic question by an experimental one: what stimulus property maximally
drives a single optic-nerve fiber, and which other stimulus variations leave
that response almost unchanged?\cite{Lettvin1959} Figure~\ref{fig:timeline}
places these works on a selective timeline. The collected volume
\textit{Embodiments of Mind} later made explicit the unity McCulloch saw among
logic, neurophysiology, and epistemology.\cite{McCulloch1965,Abraham2016}

\section{The 1943 model: a causal calculus in discrete time}
\label{sec:logicalmodel}

\subsection{Degrees of freedom and update law}

Let a directed network have $N$ units. Unit $i$ has state
$x_i(t)\in\{0,1\}$ at integer time $t$. A one denotes the occurrence of a
neural impulse during the relevant time bin. Partition the presynaptic units of
$i$ into an excitatory set $E_i$ and an inhibitory set $I_i$. A compact form of
the original-style rule is
\begin{equation}
x_i(t+1)=
\one\!\left[\sum_{j\in E_i}x_j(t)\geq\theta_i\right]
\prod_{k\in I_i}\bigl[1-x_k(t)\bigr].
\label{eq:mporiginal}
\end{equation}
The first factor implements thresholded excitatory summation. The second is an
absolute veto: if any inhibitory input is active, the product vanishes. This
distinction is easy to erase when the model is rewritten as the familiar
weighted threshold element
\begin{equation}
x_i(t+1)=\one\!\left[\sum_j w_{ij}x_j(t)-\theta_i\geq0\right].
\label{eq:weightedthreshold}
\end{equation}
Equation~\eqref{eq:weightedthreshold} is a useful later generalization. It is not
identical to Eq.~\eqref{eq:mporiginal} for arbitrary input combinations unless
negative weights are chosen sufficiently large to reproduce the veto on the
specified domain.

The unit time step represents synaptic and conduction delay after idealization.
Therefore $x_i(t+1)$ is a proposition about earlier activity, not an
instantaneous algebraic function. This gives circuit depth a temporal meaning.
It also prevents algebraic loops: even when the graph has a directed cycle, the
state at $t+1$ is determined from the state at $t$.

It helps to read Eq.~\eqref{eq:mporiginal} from right to left as a two-stage
physical test. The product first asks whether every inhibitory line is silent.
Only on that permitted branch does the threshold ask whether excitation is
sufficient. The output is then deposited one delay later. The equation thus
combines logic, causal ordering, and a specific idealization of inhibition in a
single line.

The model deliberately selects an equivalence class rather than a detailed
biophysical reconstruction. McCulloch and Pitts argued that several alternative
assumptions about threshold, delay, and inhibition can be transformed into nets
with the same input-output behavior, perhaps with a different number of units
or elapsed time.\cite{McCullochPitts1943} The invariant is the realized causal
relation, not the literal identity of every cell with one formal element.

\subsection{Propositions and adequate stimuli}

Associate the proposition $P_i(t)$ with ``unit $i$ fires at time $t$.'' Then
Eq.~\eqref{eq:mporiginal} says
\begin{equation}
P_i(t+1)\Longleftrightarrow
\left[\sum_{j\in E_i}P_j(t)\geq\theta_i\right]
\wedge\bigwedge_{k\in I_i}\neg P_k(t).
\label{eq:proposition}
\end{equation}
The square bracket denotes a cardinality proposition: at least $\theta_i$ of the
listed excitatory propositions are true. A neuron can therefore be described by
the class of preceding events sufficient for its firing. McCulloch called the
corresponding physical condition an adequate stimulus. This is a relational
claim: a spike proposes something about its causal antecedents under a model of
the intervening net.

The word ``proposition'' should not be inflated into a theory of conscious
sentences. Equation~\eqref{eq:proposition} is a formal semantics assigned by an
analyst. Multiple physical nets can realize the same proposition, and the same
physical unit can participate in different larger relations. Logical
describability alone does not settle the representational ontology of the
animal.

\section{Constructing logic from thresholded excitation and veto}
\label{sec:gates}

\subsection{AND, OR, and NOT}

Let $a,b\in\{0,1\}$. An excitatory unit with inputs $(a,b)$ and threshold two
computes
\begin{equation}
y_{\mathrm{AND}}(t+1)=\one[a(t)+b(t)\geq2]=a(t)\wedge b(t).
\end{equation}
Threshold one computes OR. NOT requires a tonic excitatory source $q(t)=1$ and
an inhibitory input from $a$:
\begin{equation}
y_{\mathrm{NOT}}(t+1)=q(t)\,[1-a(t)]=\neg a(t).
\end{equation}
Once AND and NOT are available, the network is functionally complete. Every
Boolean function has, for example, a disjunctive normal form
\begin{equation}
f(\bm{x})=\bigvee_{\bm{a}:f(\bm{a})=1}
\left(\bigwedge_{j:a_j=1}x_j\right)
\wedge
\left(\bigwedge_{j:a_j=0}\neg x_j\right).
\label{eq:dnf}
\end{equation}
Each minterm can be built by a thresholded conjunction and the minterms joined
by OR. This proves realizability, not efficiency. The number of minterms in
Eq.~\eqref{eq:dnf} can be exponential in the number of inputs, and nothing in
the construction says how a biological system should learn the circuit.
To parse the formula, choose one assignment $\bm{a}$ for which the desired
output is one. The inner conjunction recognizes exactly that row of the truth
table; the outer disjunction accepts any recognized positive row. The proof is
constructive because each syntactic layer maps directly to a circuit layer.

\subsection{Why XOR needs a network}

\begin{figure*}[t]
\centering
\includegraphics[width=.98\textwidth]{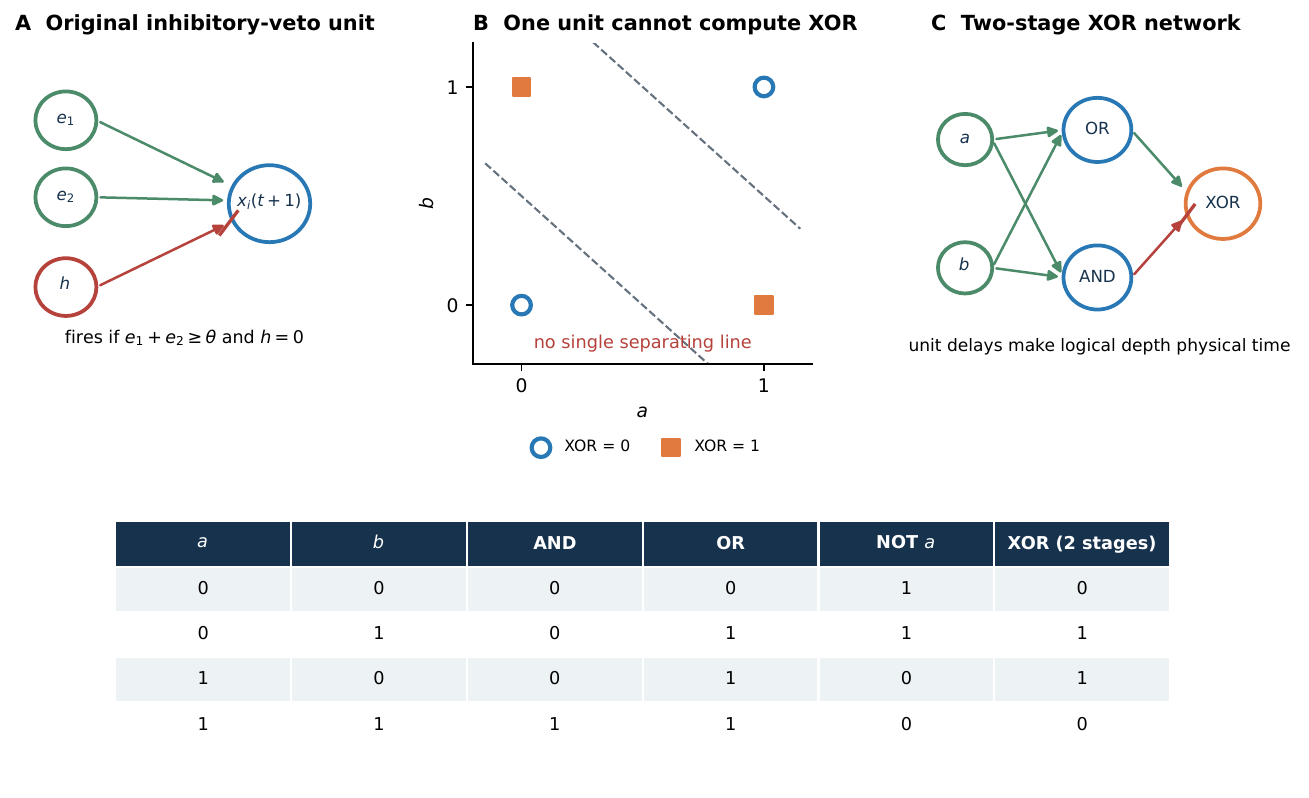}
\caption{Threshold logic. (a) In the original-style rule, excitation is summed
to threshold and any active inhibitory input vetoes firing. (b) XOR is not
linearly separable and cannot be computed by one weighted threshold unit. (c) A
two-stage construction computes XOR exactly. Each edge crossing a logical stage
also crosses a unit delay. Truth tables are generated by the supplied code.}
\label{fig:logic}
\end{figure*}

For a weighted threshold unit to compute XOR, the positive cases $(1,0)$ and
$(0,1)$ require
\begin{equation}
w_1\geq\theta,\qquad w_2\geq\theta.
\end{equation}
The negative case $(0,0)$ requires $\theta>0$, while the negative case $(1,1)$
requires $w_1+w_2<\theta$. But the first two inequalities imply
$w_1+w_2\geq2\theta>\theta$, a contradiction. XOR is not linearly separable.
It can nevertheless be constructed as
\begin{equation}
a\oplus b=(a\vee b)\wedge\neg(a\wedge b),
\label{eq:xor}
\end{equation}
using two logical stages. Figure~\ref{fig:logic} verifies the truth tables and
shows how the inhibitory veto implements the final exclusion.

\subsection{Logical depth, latency, and fan-out}

In a combinational Boolean diagram, wire delay is often ignored. In a nervous
net it cannot be. If a path from input to output contains $d$ successive units,
its result appears after $d$ time steps. Two branches that recombine must be
delay-matched if their propositions are intended to refer to the same input
time. Extra relay units can align them. Thus the equivalence of two logical
expressions does not imply equal neural latency or cost.

The abstraction also assumes reliable fan-out: the impulse of one unit may feed
many later units. Fan-out changes wiring cost but not Boolean realizability.
Conversely, restricting fan-in or requiring redundancy changes circuit size.
These distinctions foreshadow later circuit complexity, but the 1943 result is
primarily extensional: which causal propositions can a net realize?

\section{Feedback: memory, circles, and finite-state dynamics}
\label{sec:feedback}

Let $\bm{x}(t)=(x_1(t),\ldots,x_N(t))$. Any fixed synchronous net defines a map
\begin{equation}
\bm{x}(t+1)=F\bigl(\bm{x}(t),\bm{u}(t)\bigr),
\label{eq:forcedmap}
\end{equation}
where $\bm{u}(t)$ denotes peripheral input. If the input is held fixed or
removed, this becomes an autonomous map on the finite set $\{0,1\}^N$.
Directed circles make the present state depend on arbitrarily remote past input
through the current internal state. A reverberating loop can therefore retain
that an event occurred after the event itself has disappeared.

\subsection{Eventually periodic behavior}

For an autonomous deterministic net, the sequence
\begin{equation}
\bm{x}(0),\bm{x}(1),\ldots
\end{equation}
contains at most $2^N$ distinct states. By the pigeonhole principle, there exist
$0\leq t_1<t_2\leq2^N$ such that $\bm{x}(t_1)=\bm{x}(t_2)$. Determinism then
implies
\begin{equation}
\bm{x}(t_1+k)=\bm{x}(t_2+k),\qquad k\geq0.
\end{equation}
Every trajectory therefore consists of a transient of length $\mu$ followed by
a periodic orbit of length $\lambda$, with
\begin{equation}
\mu+\lambda\leq2^N.
\label{eq:eventualperiod}
\end{equation}
Fixed points are the special case $\lambda=1$. Unlike a symmetric asynchronous
Hopfield net, nothing forces $\lambda$ to equal one. Cycles are not errors in the
proof; they are part of the computational resource.

\begin{figure*}[t]
\centering
\includegraphics[width=.98\textwidth]{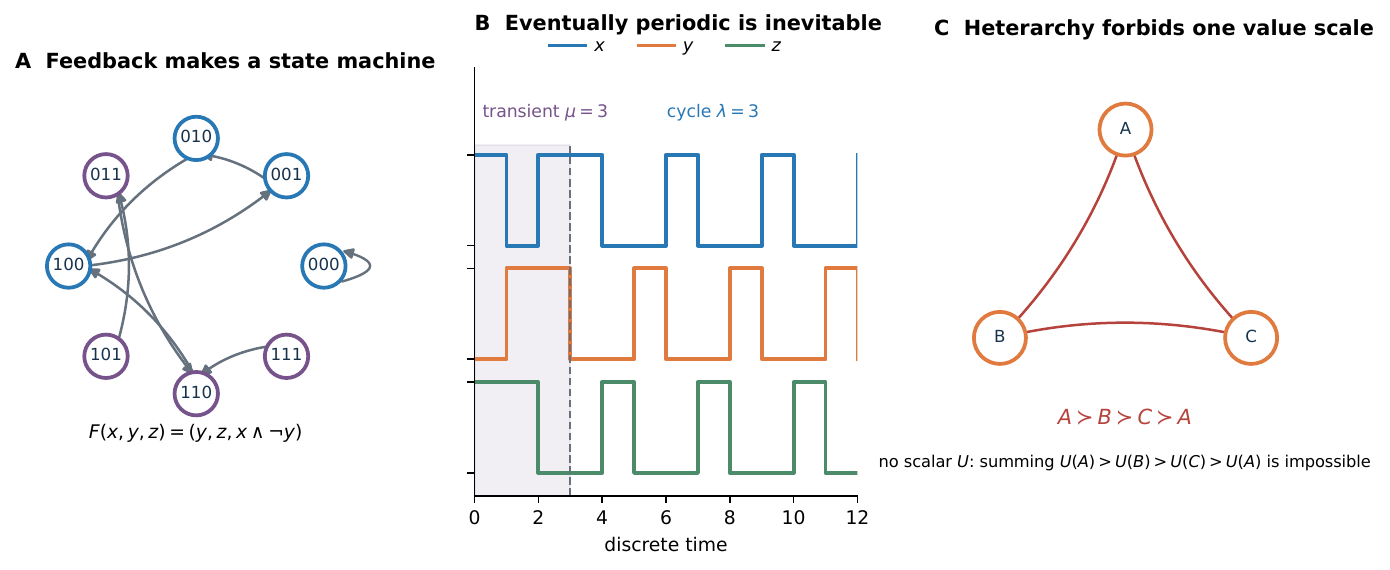}
\caption{Circularity and topology. (a) Exact state-transition graph for the
three-bit map in Eq.~\eqref{eq:ringmap}; blue nodes lie on attractors and purple
nodes are transient. (b) From $101$, a length-three transient feeds a
period-three orbit. Finite autonomous nets are eventually periodic, but need
not be invertible or begin on an attractor. (c) Cyclic preference cannot be
represented by one scalar value function. Original computational demonstrations.}
\label{fig:feedback}
\end{figure*}

Figure~\ref{fig:feedback} uses the three-bit map
\begin{equation}
F(x,y,z)=(y,z,x\wedge\neg y).
\label{eq:ringmap}
\end{equation}
Its eight states contain a fixed point, a three-cycle, and transient branches
that flow into those attractors. Starting from $101$, the trajectory reaches
the three-cycle after $\mu=3$ updates and then repeats with $\lambda=3$. The
example therefore displays both parts of Eq.~\eqref{eq:eventualperiod}:
information about the initial condition can be lost during the transient,
whereas the attractor retains only a repeating phase. Feedback changes the
object being computed from a static truth table to a temporal state machine
with basins and recurrent classes.

\subsection{Memory is state dependence, not necessarily synaptic change}

In Eq.~\eqref{eq:forcedmap}, two systems receiving the same present input can
respond differently because their internal states encode different histories.
That is dynamical memory. It does not require changing a synaptic parameter.
McCulloch distinguished this reverberatory possibility from more durable
structural memory, and the Hixon discussion made the distinction explicit.
\cite{McCulloch1951} In modern notation, state and parameter are different
carriers:
\begin{equation}
\underbrace{\bm{x}(t)}_{\text{fast dynamical memory}},\qquad
\underbrace{W(t)}_{\text{slower structural memory}}.
\end{equation}
The 1943 calculus specifies the first with fixed connections. It does not give a
general plasticity rule for the second.

\section{What ``computational universality'' does and does not mean}
\label{sec:power}

The historical literature connects McCulloch--Pitts nets to automata and
computation, especially through Kleene's later formalization.
\cite{Kleene1956} Three claims must be separated.

First, networks of threshold/veto units can realize any finite Boolean
function. Equation~\eqref{eq:dnf} proves this. Second, a recurrent network with
$N$ binary units and fixed inputs is a finite automaton with at most $2^N$
internal states. It can recognize regular temporal patterns when supplied with
an input stream. Third, a Turing machine has an unbounded tape and therefore an
unbounded number of configurations. A single fixed finite McCulloch--Pitts net
does not acquire that resource merely by containing a directed cycle.

One can simulate any finite portion of a computation by building a sufficiently
large net, or embed a controller in a system supplied with unbounded external
memory. One can also define a family of nets indexed by input size. Those are
meaningful universality statements. The unqualified slogan ``the 1943 neuron is
Turing complete'' is not.

The same distinction separates representation from learning. The existence of
a circuit for $f$ does not provide an algorithm that discovers the circuit from
examples. Rosenblatt's 1958 paper described reinforcement-style adaptation in
probabilistic perceptron models. The familiar error-correction procedure and
its finite-step convergence for linearly separable data were made explicit in
the subsequent perceptron literature, including Novikoff's theorem and
Rosenblatt's \textit{Principles of Neurodynamics}.
\cite{Rosenblatt1958,Novikoff1962,Rosenblatt1962} Minsky and Papert then
analyzed important representational limits of single-layer perceptrons; they
did not supply multilayer credit assignment.\cite{MinskyPapert1969} Training
hidden units by propagated error belongs to a later history, canonically
represented by Rumelhart, Hinton, and Williams.
\cite{RumelhartHintonWilliams1986} The sequence is therefore: realizability,
learning for a restricted architecture, analysis of that architecture's
limits, and only later a practical rule for distributing error through
multiple layers.

\subsection{Robustness is a separate theorem}

Equation~\eqref{eq:weightedthreshold} has a classification margin for a labeled
set $\{(\bm{x}^\mu,y^\mu)\}$ with $y^\mu\in\{-1,+1\}$:
\begin{equation}
\gamma=\min_\mu
\frac{y^\mu(\bm{w}^{\mathsf T}\bm{x}^\mu-b)}{\|\bm{w}\|_2}.
\label{eq:margin}
\end{equation}
If $\gamma>0$, perturbations smaller than the margin preserve the outputs on
that finite set. The 1943 equivalence arguments and later work on reliable
organisms from unreliable components ask related but distinct questions.
\cite{vonNeumann1956} Logical realizability does not automatically imply noise
tolerance, graceful degradation, or biological reliability. Those properties
require redundancy, margins, error correction, or probabilistic analysis.

\section{Heterarchy: when topology forbids a single value scale}
\label{sec:heterarchy}

McCulloch's short 1945 paper connects recurrent nervous topology to cyclic
preference. Suppose $a\succ b$ means that option $a$ is chosen over $b$ under
fixed conditions. A scalar utility representation requires a function
$U:V\to\RR$ such that
\begin{equation}
a\succ b\Longrightarrow U(a)>U(b).
\label{eq:utility}
\end{equation}
For the cycle $A\succ B$, $B\succ C$, and $C\succ A$, Eq.~\eqref{eq:utility}
would require
\begin{equation}
U(A)>U(B)>U(C)>U(A),
\end{equation}
which is impossible.

More generally, construct a directed graph with an edge $b\to a$ whenever
$a\succ b$. A finite directed graph admits a strictly increasing scalar
potential along every edge if and only if it is acyclic. The forward direction
follows because summing strict increases around a directed cycle gives a
contradiction. The reverse direction follows by topological sorting: assign
$U(v)$ equal to the vertex's rank in a topological order. Thus heterarchy is not
vagueness or lack of organization. It is organization whose directed cycles
cannot be collapsed onto one ordered line.

McCulloch's paper used the topology of intersecting dromes and the embedding of
the resulting net on surfaces. His claim was stronger in flavor than the simple
utility theorem: apparently inconsistent pairwise choices may reflect a
consistent higher-order circuit organization rather than random error.
\cite{McCulloch1945} The modern graph statement isolates the durable core
without pretending that a three-cycle by itself explains biological value.

This section also marks a decisive difference from gradient dynamics. If
behavior were always descent of one scalar $E(\bm{x})$, a strict cycle would be
impossible because $E$ would have to decrease and return to its starting value.
Heterarchical dynamics is generically nonequilibrium: local transitions need not
be integrable into a global potential.

\section{The 1947 problem of universals}
\label{sec:universals}

An organism rarely encounters the same sensory array twice. A chord can be
transposed, a shape translated or resized, and an object moved to a new retinal
location. Pitts and McCulloch asked how a nervous net could respond to a form
while disregarding such transformations. They proposed two broad mechanisms:
one combines responses over transformed presentations; the other uses feedback
to drive the presentation toward a standard form.\cite{PittsMcCulloch1947}

\subsection{Group actions and invariance}

Let a group $G$ act on a stimulus space $\mathcal{X}$. For $g\in G$ and
$x\in\mathcal{X}$, write the transformed stimulus as $g\cdot x$. A
representation $\Phi$ is invariant if
\begin{equation}
\Phi(g\cdot x)=\Phi(x),\qquad g\in G.
\label{eq:invariance}
\end{equation}
A finite-group average of a feature $\phi$ is
\begin{equation}
\Phi_G(x)=\frac{1}{|G|}\sum_{g\in G}\phi(g\cdot x).
\label{eq:groupaverage}
\end{equation}
For any $h\in G$,
\begin{align}
\Phi_G(h\cdot x)
&=\frac{1}{|G|}\sum_{g\in G}\phi(g h\cdot x)\\
&=\frac{1}{|G|}\sum_{g'\in G}\phi(g'\cdot x)=\Phi_G(x),
\end{align}
because $g\mapsto gh$ permutes the group. For a compact continuous group, the
sum is replaced by integration against normalized Haar measure.

Equation~\eqref{eq:groupaverage} is modern notation for the paper's averaging
logic, not a claim that the authors wrote current group-equivariant network
formalism. It exposes the physics: symmetry identifies an orbit
\begin{equation}
\mathcal{O}_x=\{g\cdot x:g\in G\},
\end{equation}
and an invariant representation assigns the same value to every point on that
orbit.
The average performs two conceptual operations at once. It first generates all
presentations declared equivalent by $G$, then forgets which presentation was
encountered by summing over the orbit. The proof works because transforming the
input merely reorders the same summands. That simplicity is also the danger:
the averaging step can discard information that the task did not intend to
treat as nuisance.

\subsection{Invariant, equivariant, and selective}

Invariance is useful only to nuisance transformations. If all spatial
information is averaged away at the first stage, distinct objects can become
indistinguishable. An equivariant feature map instead satisfies
\begin{equation}
\Psi(g\cdot x)=\rho(g)\Psi(x),
\label{eq:equivariance}
\end{equation}
where $\rho(g)$ transforms the feature coordinates. A retinotopic map is
approximately translation equivariant: moving a stimulus moves the activity
pattern. Pooling later can produce invariance while preserving selectivity in
earlier layers.

For template matching, one useful orbit score is
\begin{equation}
S(x)=\max_{g\in G}
\frac{\langle x,g\cdot\xi\rangle}
{\|x\|_2\,\|\xi\|_2},
\label{eq:orbitmax}
\end{equation}
where $\xi$ is a reference form. If the searched transformations cover the
nuisance orbit, $S$ remains high when $x$ is shifted or rescaled. The price is
computation over the orbit and possible false matches when different orbits
overlap.

\begin{figure*}[t]
\centering
\includegraphics[width=.98\textwidth]{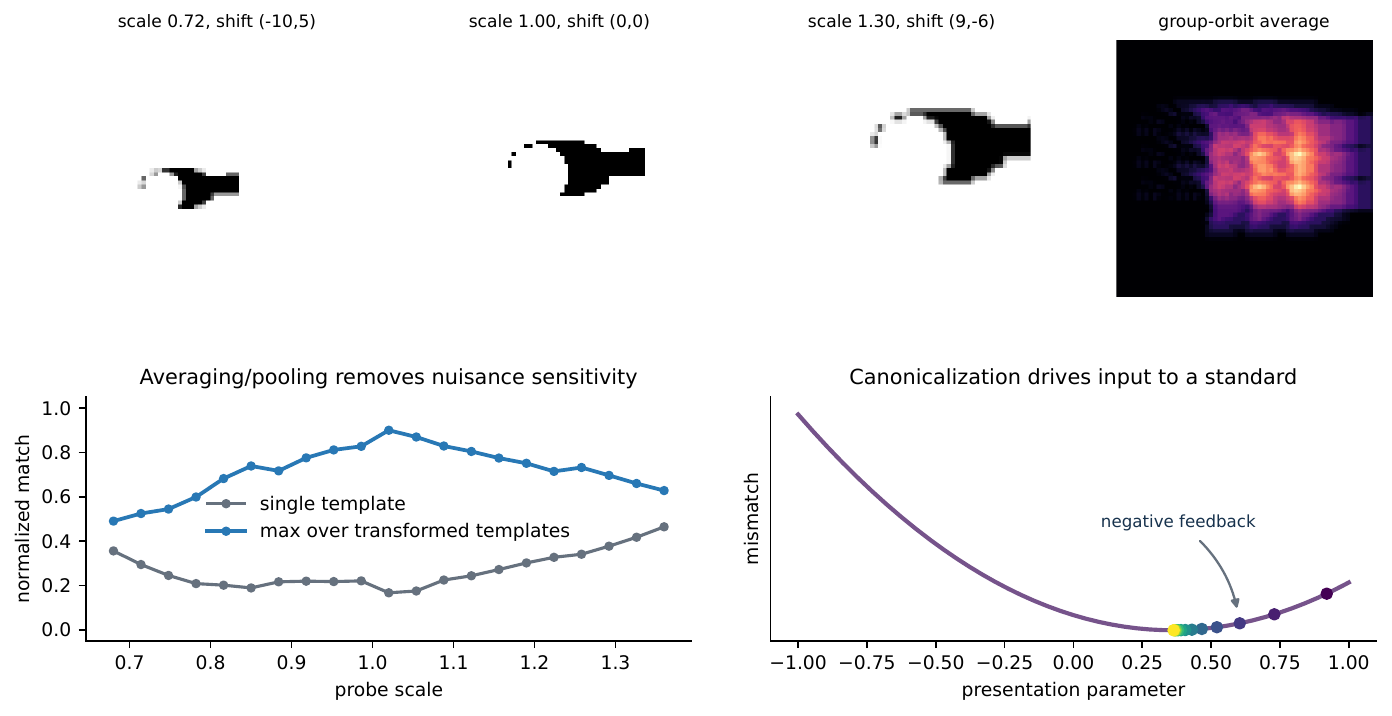}
\caption{Two routes to a universal. Top: translated and rescaled presentations
belong to one transformation orbit; their average is invariant but blurred.
Bottom left: max matching over an orbit reduces nuisance sensitivity relative to
one template. Bottom right: negative feedback drives a presentation parameter
toward a canonical value, reducing mismatch. These are modern computational
realizations of the two mechanism classes in Pitts and McCulloch (1947), not
reconstructions of their exact diagrams.}
\label{fig:universals}
\end{figure*}

Figure~\ref{fig:universals} compares one fixed template with max pooling over 45
translated and scaled templates. The mean normalized match rises from 0.270 to
0.717 over the tested probes. This is a finite demonstration of the mechanism,
not a measurement from the 1947 paper. Related ideas later appear in
position-tolerant recognition, the neocognitron, and explicitly group-
equivariant networks.\cite{Fukushima1980,CohenWelling2016}

\section{Canonicalization by feedback}
\label{sec:canonical}

The second route to invariance estimates the transformation and reverses it.
Let $\widehat g(x)$ be an estimated pose, scale, pitch shift, or retinal
displacement. A canonicalized representation is
\begin{equation}
C(x)=\widehat g(x)^{-1}\cdot x.
\label{eq:canonicalization}
\end{equation}
If $\widehat g(h\cdot x)=h\widehat g(x)$, then
$C(h\cdot x)=C(x)$. Biological orienting movements can participate in such a
computation: the organism changes the sensory presentation rather than
constructing every transformed template internally.

A simple feedback model adjusts a presentation coordinate $a$ to minimize
mismatch $L(a)$:
\begin{equation}
\tau\dot a=-\kappa\frac{\partial L}{\partial a},\qquad \kappa>0.
\label{eq:feedbackcanon}
\end{equation}
Along the trajectory,
\begin{equation}
\frac{\dd L}{\dd t}=\frac{\partial L}{\partial a}\dot a
=-\frac{\kappa}{\tau}\left(\frac{\partial L}{\partial a}\right)^2\leq0.
\label{eq:canonLyapunov}
\end{equation}
The mismatch is a Lyapunov function for this one-dimensional idealization. In
the coded example, the absolute pose error falls from 0.56 to 0.0038 after
twelve discrete corrections.
Equation~\eqref{eq:canonLyapunov} is the bridge from an abstract invariance
claim to dynamics. The first factor says how mismatch changes with pose; the
feedback law chooses motion in exactly the opposite direction. Their product
is therefore nonpositive. A canonical view is reached by dissipating mismatch,
not by averaging away pose.

Canonicalization has failure modes that averaging hides. Symmetric objects can
have multiple equally good canonical poses. An estimator can jump
discontinuously when the stimulus crosses an ambiguity. Feedback can become
unstable under delay or excessive gain. These are physical constraints on an
apparently abstract invariant.

The 1947 paper is therefore not merely a precursor to pooling. Its deeper claim
is that perception of a universal can be implemented either by internal
convergence across transformed instances or by a closed sensorimotor loop that
standardizes the instance. In both cases, invariance is constructed by dynamics
and anatomy; it is not a label attached after the fact.

\section{From synthetic nets to the frog's visual system}
\label{sec:frog}

\subsection{The experimental question}

``What the Frog's Eye Tells the Frog's Brain'' begins with the animal. A frog
hunts and evades in a visual world; its stabilized eyes do not scan in the same
way as primate eyes. The authors then ask what single optic-nerve fibers report.
Their method was to search broadly for a stimulus that produced strong activity
in one fiber and to vary other aspects of the stimulus to determine which
property was essential.\cite{Lettvin1959}

This approach differs from presenting only spots or full-field flashes and
assuming that all interpretation occurs later. The authors used spots, edges,
geometric objects, movement, and behaviorally suggestive targets. They recorded
from single myelinated and unmyelinated fibers in the intact optic nerve of
\textit{Rana pipiens}, and also from terminal activity in superficial tectal
layers. The experiment was constrained by prior work on optic-nerve receptive
fields, retinal summation and inhibition, and center-surround organization.
\cite{Hartline1938,Hartline1940,Barlow1953,Kuffler1953}

The retina contained many more receptors than optic-nerve fibers, and each
ganglion cell pooled from many receptors. A pointwise luminance copy was
therefore already implausible. The main finding was a decomposition into four
principal parallel operations, each distributed across the retina and nearly
independent of general illumination. The paper also briefly reported a small,
rarer fifth group without distinct receptive fields whose discharge tracked
absolute darkness over a wide area.\cite{Lettvin1959}

\begin{table*}[t]
\caption{The four principal operations reported in the 1959 frog paper. The
authors additionally noted a rare wide-field absolute-darkness group; it is not
expanded into a fifth principal channel because their discussion and tectal
organization centered on the four operations tabulated here. Later retinal
taxonomy is richer still.}
\label{tab:frogops}
\begin{ruledtabular}
\begin{tabular}{p{.18\textwidth}p{.34\textwidth}p{.38\textwidth}}
Operation & Strong stimulus property & Important qualifications \\ \hline
Sustained contrast & A sharp light-dark boundary, moving or stationary & Small receptive field; little response to uniform on/off by itself; activity can persist while the boundary remains.\\
Net convexity & A small dark object or positively curved dark boundary within the field & Straight edges are ineffective; intermittent motion is especially effective; transient darkness can erase the sustained response.\\
Moving edge & A boundary moving through a larger receptive field & Related to ON-OFF fibers; response grows with velocity over a range and changes little across large general-illumination changes.\\
Net dimming & Rapid decrease of illumination over the largest field & Fast myelinated afferents; prolonged, sometimes synchronized discharge after sudden darkening.\\
\end{tabular}
\end{ruledtabular}
\end{table*}

\subsection{Parallel maps, not four labels at one place}

Each operation was represented across retinal position. The optic-nerve fibers
were braided rather than spatially ordered within the nerve, but their terminals
formed four depth-separated, registered retinotopic maps in the tectum. Contrast
terminals were most superficial, convexity terminals below them, moving-edge
terminals deeper, and dimming terminals deepest. The 1959 paper states both the
braiding and the orderly tectal maps; the 1960 anatomical-physiological companion
develops this organization in greater detail.\cite{Lettvin1959,Maturana1960} This combines
two organizations:
\begin{equation}
\text{channel identity}\times\text{retinal position}.
\end{equation}
The output is neither a raw camera image nor a four-dimensional global label.
It is a set of feature-specific spatial maps.

The authors sometimes used the memorable phrase ``bug perceiver'' for a net-
convexity fiber. It is pedagogically dangerous when detached from the actual
response conditions. The unit did not semantically recognize the biological
category ``bug'' in every context. It responded to a conjunction of size,
contrast polarity, boundary geometry, position, and motion history that is
useful for accessible prey. Behavioral relevance motivates the stimulus family;
it does not turn one axon into a proposition with a context-free dictionary
definition.

\subsection{What the paper changed}

The experiment moved visual computation into the retina. Local variation,
boundary geometry, motion, and dimming were already separated before the signal
reached deeper brain structures. This was contemporaneous with Hubel and
Wiesel's early cortical receptive-field work and belongs to a broader shift from
point illumination to feature-selective transformations.
\cite{HubelWiesel1959,Barlow1961} The specifically McCullochian continuity is
that a spike is read through its adequate stimulus: what invariant property of
the preceding sensory world does the fiber's activity propose?

The paper also states its scope unusually clearly: the interpretation applies to
the frog. Retinal cell types, ecological demands, and downstream circuits vary
across species. Modern retinal physiology reveals many more computations and
adaptive cell classes than four.\cite{GollischMeister2010,Donner2020} The 1959
result is foundational because of its experimental logic and parallel feature
decomposition. Even within the original article, the rare absolute-darkness
fibers already warn against mistaking the four principal operations for an
exhaustive atlas.

\section{An effective model of the four retinal operations}
\label{sec:retinamodel}

The original paper inferred operations from recordings but did not supply one
complete quantitative circuit model for all four. We therefore construct a
minimal effective model whose purpose is to make the selectivities inspectable.
It is not a fit to the historical traces.

Let $I(x,t)>0$ be image intensity along one retinal coordinate. Define local and
broad Gaussian averages $G_\sigma*I$ and $G_\Sigma*I$, with $\Sigma>\sigma$.
A divisively normalized contrast field is
\begin{equation}
c(x,t)=\frac{(G_\sigma-G_\Sigma)*I(x,t)}
{\epsilon+G_\Sigma*I(x,t)}.
\label{eq:normcontrast}
\end{equation}
For a multiplicative illumination change $I\mapsto\alpha I$, $c$ is exactly
invariant when $\epsilon=0$ and approximately invariant when the broad signal
dominates $\epsilon$. This is one mathematical route to the paper's observation
that responses changed little across large illumination variation. Divisive
normalization is a general neural computation, but Eq.~\eqref{eq:normcontrast}
should not be attributed as the authors' literal 1959 circuit.
\cite{CarandiniHeeger2012}

With a rectifier $[z]_+=\max(z,0)$, define the qualitative fields
\begin{align}
q_{\mathrm{con}}(x,t)&=|c(x,t)|,\label{eq:qcontrast}\\
d(x,t)&=[-c(x,t)]_+,\\
q_{\mathrm{conv}}(x,t)&=[d(x,t)-\lambda(G_\rho*d)(x,t)]_+,
\label{eq:qconvex}\\
q_{\mathrm{mov}}(x,t)&=|\partial_t c(x,t)|,
\label{eq:qmoving}\\
q_{\mathrm{dim}}(x,t)&=
\left[-\partial_t\log\bigl((G_\Sigma*I)(x,t)+\epsilon\bigr)\right]_+.
\label{eq:qdimming}
\end{align}
The second line is a compact-dark-object proxy: local darkness excites while
broad dark occupancy suppresses. It should be called a convexity proxy, because
a complete curvature detector would require orientation-resolved boundary
signals and a more explicit geometric circuit. The moving-edge channel detects
temporal change in normalized contrast; the dimming channel detects a negative
log-intensity derivative, which is invariant to constant multiplicative scale.
Here $\epsilon>0$ is only a numerical floor \emph{inside} the logarithm. It is
not an additive firing bias outside the derivative, so a static image produces
$q_{\mathrm{dim}}=0$ rather than a positive baseline. This is the same operation
implemented by the supplied code, whose simulated intensity is strictly
positive.

\begin{figure*}[t]
\centering
\includegraphics[width=.98\textwidth]{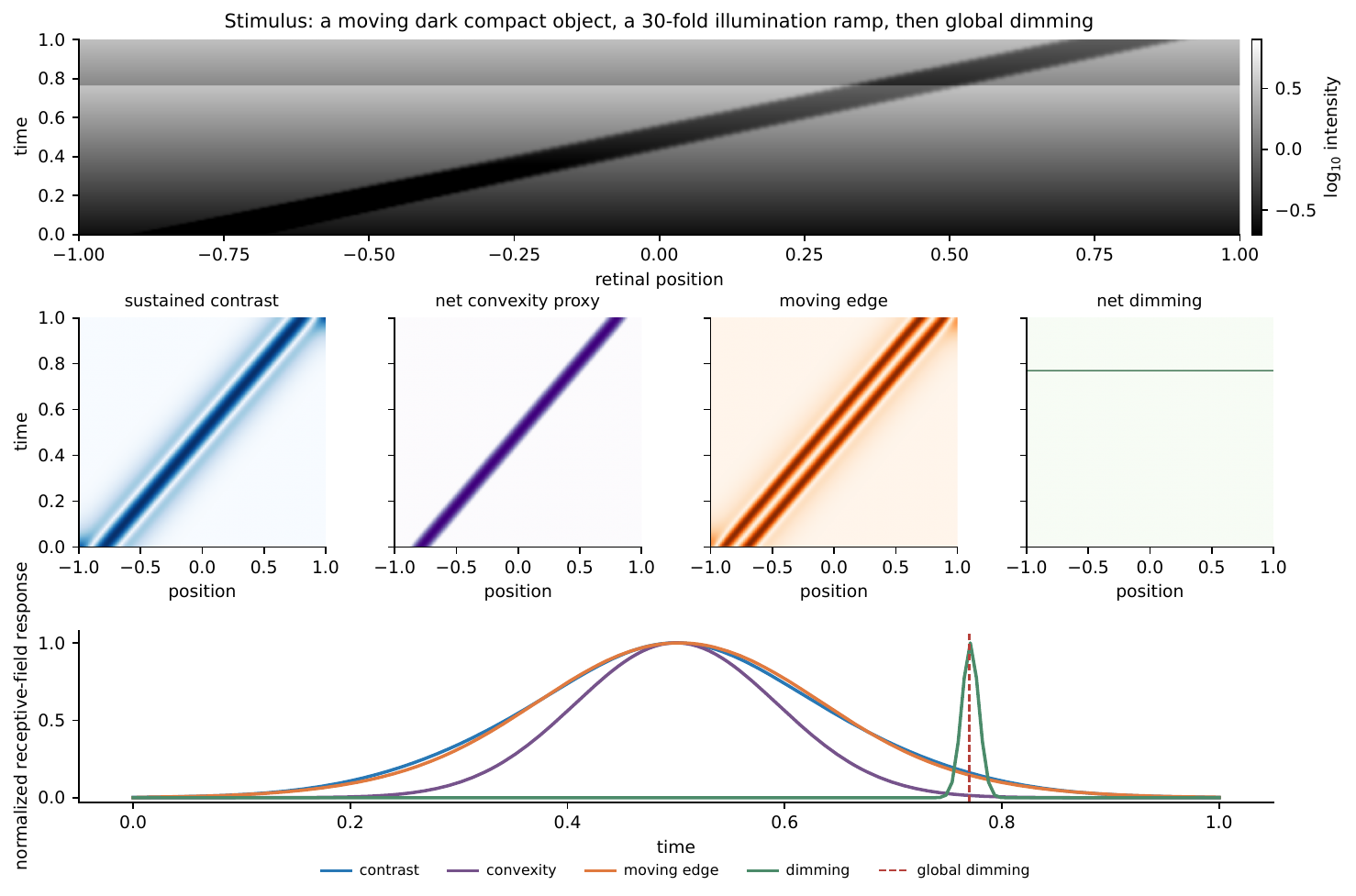}
\caption{A qualitative effective model of the four frog-retina operations. The
stimulus is a moving dark compact object during a 30-fold illumination ramp,
followed by global dimming. Normalized spatial and temporal filters create
parallel contrast, compact-dark-object, moving-edge, and dimming fields. Bottom:
readout from a representative central receptive field. This is a pedagogical
LN/LNP-style construction, not a fit or digitization of the 1959 data.}
\label{fig:frogmodel}
\end{figure*}

Figure~\ref{fig:frogmodel} drives these fields with a dark compact object moving
across the retina during an approximately 30-fold illumination ramp, followed
by a sudden global dimming. Contrast, compact darkness, and motion follow the
object across position. The dimming field remains quiet until the global step.
A representative receptive field responds when the object crosses its center,
while the dimming readout peaks at the step. The code saves every intermediate
field so readers can change widths, gains, and time constants.

\subsection{From filter field to spikes}

A linear-nonlinear-Poisson description converts a stimulus history
$\bm{s}_t$ into a conditional rate
\begin{equation}
\lambda(t)=f\bigl(\bm{k}^{\mathsf T}\bm{s}_t+b\bigr),
\label{eq:lnp}
\end{equation}
and, for a short bin $\Delta t$,
\begin{equation}
\Prob[n_t=1\mid\bm{s}_t]=1-e^{-\lambda(t)\Delta t}.
\label{eq:poissonbin}
\end{equation}
Here $\bm{k}$ is a spatiotemporal filter and $f$ includes rectification,
saturation, and adaptation. A single linear filter is insufficient for
convexity or contrast energy, which depend on multiple subunits or quadratic
combinations. The hierarchy
\begin{equation}
\begin{aligned}
\text{linear filter}&\rightarrow\text{nonlinearity}\\
&\rightarrow\text{normalization/adaptation}\rightarrow\text{spikes}
\end{aligned}
\end{equation}
turns verbal response classes into falsifiable models.

\section{Adequate stimuli as a system-identification problem}
\label{sec:identification}

The frog paper's method can be stated as controlled optimization over stimulus
space. Let $R[s]$ be a response functional. One seeks a stimulus $s^*$ that
drives the fiber strongly,
\begin{equation}
s^*\in\mathop{\mathrm{arg\,max}}_{s\in\mathcal{S}}\E[R\mid s],
\label{eq:argmaxstim}
\end{equation}
then changes candidate nuisance variables while preserving the proposed
feature. A feature hypothesis is strong when it predicts both excitation and
invariance. This is richer than finding one dramatic stimulus: many different
models can share one maximizer but disagree away from it.

\subsection{Spike-triggered averaging}

For zero-mean Gaussian white stimulus vectors $\bm{s}$, the spike-triggered
average is
\begin{equation}
\mathrm{STA}=\E[\bm{s}\mid\text{spike}]-\E[\bm{s}].
\label{eq:sta}
\end{equation}
If the firing probability depends monotonically on the linear projection
$\bm{k}^{\mathsf T}\bm{s}$, rotational symmetry of the Gaussian ensemble makes
the conditional mean parallel to $\bm{k}$:
\begin{equation}
\mathrm{STA}=a\bm{k}
\label{eq:staparallel}
\end{equation}
for a scalar $a$ determined by the nonlinearity and rate. Thus STA can recover a
linear adequate feature up to scale.

But suppose the response is an even contrast-energy function
\begin{equation}
\lambda(\bm{s})=f\bigl[(\bm{k}^{\mathsf T}\bm{s})^2\bigr].
\label{eq:energyresponse}
\end{equation}
For every spike-driving stimulus $\bm{s}$, the opposite stimulus $-\bm{s}$ has
the same rate. Their contributions to Eq.~\eqref{eq:sta} cancel, giving zero
even though the neuron is highly selective. The spike-triggered covariance
\begin{equation}
C_{\mathrm{ST}}=Cov[\bm{s}\mid\text{spike}]-\Cov[\bm{s}]
\label{eq:stc}
\end{equation}
can instead reveal a direction proportional to $\bm{k}\bm{k}^{\mathsf T}$.

\begin{figure*}[t]
\centering
\includegraphics[width=.98\textwidth]{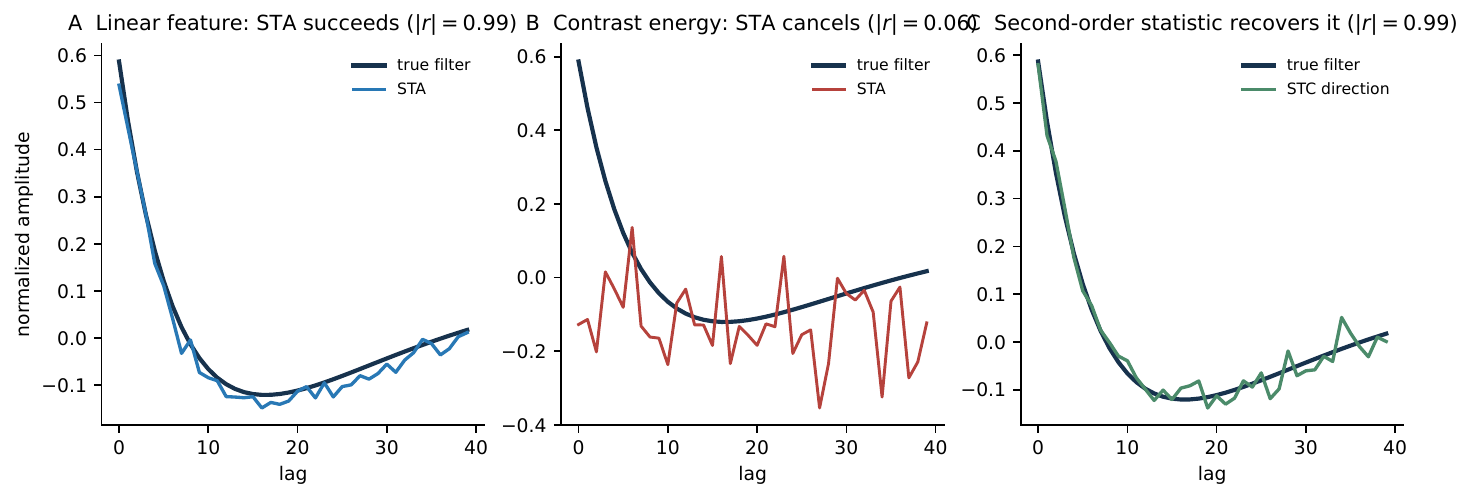}
\caption{Adequate stimuli and identifiability. (a) For a linear feature under
Gaussian stimulation, spike-triggered averaging recovers the filter. (b) For an
even contrast-energy response, opposite stimuli drive equally and the STA
cancels. (c) A second-order spike-triggered covariance direction recovers the
feature. Exact numerical values are recorded in
\texttt{numerical\_results.json}.}
\label{fig:identification}
\end{figure*}

Figure~\ref{fig:identification} uses 120,000 seeded samples. For a linear-
nonlinear unit, the recovered STA has absolute correlation 0.990 with the true
filter. For the energy unit, the STA correlation falls to 0.061, while the
leading spike-triggered covariance direction recovers the filter with
correlation 0.989. The failure of a method can therefore masquerade as absence
of selectivity. This is precisely why adequate-stimulus searches must explore
the relevant nonlinear stimulus coordinates.

The example also clarifies the frog paper's use of structured stimuli. White
noise is not automatically superior to behaviorally meaningful probes. A
stimulus ensemble determines which model dimensions are identifiable and how
efficiently they are sampled. Naturalistic objects can expose nonlinear
conjunctions quickly; controlled perturbations then test which conjunction
terms are causal. Modern analyses use maximally informative dimensions,
subunit models, generalized linear models, and closed-loop stimulus design, but
the logical sequence remains recognizable.\cite{Rieke1997,Rust2005}

\section{What exactly was McCulloch's scientific advance?}
\label{sec:advance}

The following claims summarize the durable contribution without turning one
person into the sole origin of neural networks, AI, or receptive-field theory.

\begin{enumerate}[leftmargin=*]
\item \emph{Neural events became elements of a causal formal system.} A spike
could be interpreted as a proposition about its adequate antecedent, and a net
could be analyzed as a temporally indexed logical relation.

\item \emph{The program was bidirectional.} The 1943 calculus described the
behavior of given nets and synthesized nets for admissible expressions. It
therefore joined neurophysiological modeling to circuit construction.

\item \emph{Feedback became a computational state variable.} Circles gave
reference to an indefinite past, sustained internal activity, and temporal
patterns. They also made clear that recurrent computation need not be descent
of a scalar energy.

\item \emph{Topology constrained value.} Cyclic preference was related to
network organization that cannot be embedded in a single scalar hierarchy.

\item \emph{Invariance became a mechanism to be built.} Averaging over
transformations and feedback to a canonical presentation turned the philosophical
problem of universals into candidate circuits.

\item \emph{The same adequate-stimulus logic became experimental.} The frog
work inferred what single fibers report by maximizing response and varying
irrelevant stimulus dimensions, revealing parallel feature maps before the
brain proper.
\end{enumerate}

The arc is unusual: logical calculus, topology, group-like invariance, control,
and single-fiber physiology are not separate achievements accidentally sharing
an author. They instantiate a proposed experimental epistemology. One asks what
relations a physical knower can preserve, how its organization preserves them,
and what intervention reveals that organization.\cite{Arbib2000,McCulloch1961}

\begin{table*}[t]
\caption{Tempting shorthand and the statement that should replace it.}
\label{tab:precision}
\begin{ruledtabular}
\begin{tabular}{p{.28\textwidth}p{.65\textwidth}}
Shorthand & Precise statement \\ \hline
``McCulloch invented the artificial neuron.'' & McCulloch and Pitts made a foundational synthesis of threshold neurophysiology, temporal logic, and circuit realizability within an existing mathematical-biophysics and switching-theory context.\\
``One McCulloch--Pitts neuron computes any Boolean function.'' & One weighted threshold element computes a linearly separable Boolean function. Networks built from thresholded excitation and inhibition can realize every finite Boolean function.\\
``Inhibition is just a negative weight.'' & In the original idealization, any active inhibitory input vetoes firing. A finite negative weight reproduces that rule only when its magnitude and input domain guarantee a veto.\\
``The 1943 net is Turing complete.'' & A fixed recurrent net with $N$ binary units is a finite-state machine with at most $2^N$ states. Unbounded computation requires a growing family, external memory, or another unbounded resource.\\
``The paper introduced learning in neural networks.'' & It studied fixed nets, logical realizability, and memory through circles or possible alteration; it did not provide a general data-driven weight-learning algorithm.\\
``Logical description proves that the brain manipulates symbols.'' & Neural firing can be assigned a propositional semantics relative to a model of its antecedents. That formal equivalence does not by itself establish explicit cognitive symbols.\\
``How We Know Universals invented group convolution.'' & The 1947 paper proposed averaging/scansion over transformations and feedback standardization. Group averaging and equivariant networks are useful modern formalizations, not identical historical algorithms.\\
``The frog retina contains bug detectors.'' & Net-convexity fibers responded to a behaviorally relevant conjunction involving a small dark object, curved boundary, position, motion, and history. ``Bug perceiver'' was a suggestive functional gloss.\\
``The retina does not send an image.'' & The frog optic nerve carried four principal, registered retinotopic feature maps rather than a pointwise map of luminance alone; the paper also noted rare wide-field darkness fibers. Spatial organization was transformed, not abolished.\\
``The 1959 paper discovered receptive fields.'' & Hartline, Barlow, Kuffler, and others established essential receptive-field results. Lettvin and colleagues used broader stimulus search to identify four principal operation classes, a rarer fifth group, and their tectal organization.\\
\end{tabular}
\end{ruledtabular}
\end{table*}

\section{Limits, biological interpretation, and effective scale}
\label{sec:limits}

The McCulloch--Pitts net is too idealized to be a cellular reconstruction and
too exact to serve as an unconstrained brain metaphor. Its proper role is an
effective theory of causal organization. The binary state discards spike time
within a bin, firing rate, dendritic location, adaptation, neurotransmitter,
and stochasticity. Unit delay suppresses a distribution of conduction times.
Absolute inhibition ignores graded shunting and conductance dynamics. Fixed
thresholds omit state dependence and homeostasis.

Coarse graining is justified only when the retained variables close. If two
microscopic states with the same binary $\bm{x}(t)$ systematically lead to
different $\bm{x}(t+1)$ because of adaptation or dendritic voltage, then
$\bm{x}$ is not a sufficient Markov state. One must enlarge it, perhaps to
\begin{equation}
\bm{z}(t)=\bigl(\bm{x}(t),\bm{a}(t),\bm{v}_{\mathrm{dend}}(t),\ldots\bigr).
\end{equation}
The convenience of a point-neuron binary description does not guarantee that
it is the right observational scale. Conversely, if omitted variables only
renormalize thresholds, effective delays, or noise over the task timescale, the
logical model can remain informative.

The frog channel model has analogous limits. Equations~\eqref{eq:qcontrast}--
\eqref{eq:qdimming} are operation-level approximations. A morphological model
would include photoreceptors, bipolar, horizontal, amacrine, and ganglion-cell
subtypes, synaptic rectification, adaptation, and retinal feedback. The
effective model earns its place by predicting stimulus-response relations, not
by resembling every anatomical detail.

This stance strengthens rather than diminishes McCulloch's work. The 1943 paper
explicitly cared about invariance under changes of implementation. The 1959
paper then tested whether operation-level descriptions were visible at an
actual biological output. Together they pose a general standard: choose the
coarsest scale that preserves the causal relation of interest, and test closure
rather than assuming it.

\section{Computational companion}
\label{sec:repro}

The supplied program requires NumPy, SciPy, and Matplotlib. From the project
directory, run
\begin{verbatim}
python code/mcculloch_replication.py \
  --output figures \
  --results numerical_results.json
\end{verbatim}
The fixed seed is 1943 unless changed with \texttt{--seed}. Every figure is
saved as PDF, PNG, and SVG. The JSON file records truth tables, exact state
transitions and cycles, transformation-pooling scores, canonicalization error,
retinal stimulus parameters, and system-identification correlations.

\begin{table*}[t]
\caption{What each computational figure establishes.}
\label{tab:figures}
\begin{ruledtabular}
\begin{tabular}{p{.15\textwidth}p{.78\textwidth}}
Figure & Reproduced mechanism \\ \hline
\ref{fig:logic} & veto inhibition, Boolean gates, XOR nonseparability, and multilayer construction \\
\ref{fig:feedback} & finite-state transients, attractor cycles, and obstruction to scalar utility \\
\ref{fig:universals} & orbit pooling and feedback canonicalization \\
\ref{fig:frogmodel} & four qualitative normalized feature channels under illumination change \\
\ref{fig:identification} & linear STA recovery, energy-model cancellation, and STC recovery \\
\end{tabular}
\end{ruledtabular}
\end{table*}

The numerical figures support mechanism-level claims. Figure~\ref{fig:logic}
does not establish biological sufficiency of binary threshold units.
Figure~\ref{fig:universals} does not reconstruct the exact 1947 circuits.
Figure~\ref{fig:frogmodel} does not replicate historical spike trains, because
no raw recordings are used. Figure~\ref{fig:identification} establishes an
identifiability phenomenon under a specified Gaussian ensemble, not a complete
analysis of frog ganglion cells. The primary papers supply the historical
observations; the code makes the mathematical mechanisms testable.

\section{Guided derivations and computational exercises}
\label{sec:exercises}

\subsection{Threshold logic}
\begin{enumerate}[leftmargin=*]
\item Starting from Eq.~\eqref{eq:mporiginal}, construct NAND using the smallest
network you can under the assumption that a tonic source is available. Repeat
without a tonic source and state what extra temporal input is required.
\item Find a finite negative inhibitory weight in
Eq.~\eqref{eq:weightedthreshold} that exactly reproduces one veto input when
there are $m$ unit excitatory inputs and threshold $\theta$. Show why the answer
depends on a bound on total excitation.
\item Prove XOR nonseparability with an explicit bias $b$ instead of threshold
$\theta$. Then give weights for the two hidden units and output unit in
Eq.~\eqref{eq:xor}.
\item Convert a three-variable Boolean function to disjunctive normal form and
build its delayed net. Insert relay units so every minterm arrives at the output
at the same time.
\item Count units and edges for the direct DNF construction of parity on $n$
bits. Compare it with a tree of two-input XOR subnetworks. Separate depth, size,
and fan-in.
\end{enumerate}

\subsection{Feedback and automata}
\begin{enumerate}[leftmargin=*]
\item Enumerate Eq.~\eqref{eq:ringmap} by hand. Verify the fixed point, the
three-cycle, and the $\mu=3$ transient from $101$ in
Figure~\ref{fig:feedback}. Replace the vetoed conjunction by XOR and determine
which transients disappear when the update becomes invertible.
\item Prove Eq.~\eqref{eq:eventualperiod}. Construct an $N$-bit map with one
cycle of length $2^N$ using a binary counter, and discuss the threshold-circuit
cost of its update.
\item Add a one-bit external input to the three-bit map. Build the transition
matrices $T_0$ and $T_1$. Give an input word that synchronizes at least two
initial states.
\item Design a recurrent network that reports whether the last two input bits
were 01. Minimize internal states before translating the automaton to threshold
units.
\item Explain why eventual periodicity fails as a useful bound when a system is
continuously driven by a nonperiodic input, even though its internal state set
remains finite.
\end{enumerate}

\subsection{Heterarchy and potentials}
\begin{enumerate}[leftmargin=*]
\item Prove that a finite directed graph admits a strictly increasing potential
on all edges if and only if it is acyclic.
\item For noisy pairwise choices, fit a Bradley--Terry scalar score and compute
the residual circulation around each triangle. Explain what noise can and cannot
show about structural heterarchy.
\item Construct a continuous two-dimensional vector field with a stable limit
cycle. Show that it cannot be written globally as $-\nabla E$ under the Euclidean
metric.
\item Compare a rock-paper-scissors replicator dynamic with the static
preference cycle. Which additional assumptions are required to move from a
choice graph to a dynamical system?
\end{enumerate}

\subsection{Universals and symmetry}
\begin{enumerate}[leftmargin=*]
\item Prove Eq.~\eqref{eq:invariance} for a finite group average by reindexing.
Repeat for Haar integration and state the measure-invariance property used.
\item Give two nonidentical images whose raw translation-orbit averages are the
same. Explain why invariance can destroy selectivity.
\item Show that convolution is translation equivariant. Identify which boundary
condition is required for exact equivariance on a finite array.
\item Modify the replication code to include rotations. Plot recognition score
versus the number of sampled group elements and quantify the accuracy-cost
tradeoff.
\item Let an object have a nontrivial stabilizer
$G_x=\{g:g\cdot x=x\}$. Explain why canonicalization may be nonunique and how an
equivariant representation can avoid choosing one pose.
\item Add delay to Eq.~\eqref{eq:feedbackcanon},
$\tau\dot a(t)=-\kappa L'[a(t-d)]$. Linearize near the optimum and find the
stability boundary for a quadratic $L$.
\end{enumerate}

\subsection{Frog retina and system identification}
\begin{enumerate}[leftmargin=*]
\item Prove the illumination invariance of Eq.~\eqref{eq:normcontrast} for
$\epsilon=0$. With $\epsilon>0$, derive an upper bound on the change caused by
$I\mapsto\alpha I$.
\item Replace the one-dimensional stimulus by a two-dimensional moving disk.
Construct an orientation-resolved curvature score and compare it with the
compact-dark-object proxy in Eq.~\eqref{eq:qconvex}.
\item Generate a stationary edge, a large translating bar, a small dark disk,
and a full-field dimming step. Make a response matrix for the four model
channels. Determine whether the matrix uniquely identifies them.
\item Derive Eq.~\eqref{eq:staparallel} for Gaussian stimuli using the
decomposition $\bm{s}=z\widehat{\bm{k}}+\bm{s}_\perp$. Which step fails for a
correlated stimulus ensemble?
\item Prove that the STA is zero for Eq.~\eqref{eq:energyresponse} under a
stimulus distribution symmetric under $\bm{s}\mapsto-\bm{s}$. Derive the form
of the spike-triggered covariance change.
\item Reproduce Figure~\ref{fig:identification} with correlated naturalistic
noise. Whiten the stimuli before STA/STC and compare filter recovery.
\item Design a closed-loop experiment that distinguishes a genuine curvature
detector from a small-dark-object detector while matching area, contrast,
velocity, and mean luminance.
\end{enumerate}

\section{A primary-literature reading sequence}
\label{sec:reading}

Begin with the abstract and physiological assumptions of the 1943 paper, then
translate the first simple nets into Eq.~\eqref{eq:mporiginal}. Read the paper
twice. On the first pass, mark the definitions of equivalence, realizability,
and circles. On the second, follow how temporal reference and quantified
expressions enter. Do not let difficult historical notation obscure the two
directions: net-to-expression and expression-to-net.
\cite{McCullochPitts1943}

Read Shannon and Turing beside it, not to collapse their projects, but to
separate switching algebra, effective procedure, and nervous-net synthesis.
\cite{Shannon1938,Turing1936} Then read Kleene's 1956 treatment to see how
events in nerve nets become finite automata and regular expressions.
\cite{Kleene1956} At that point, write one paragraph explaining why a fixed
finite net is not an unbounded Turing machine.

Next read the 1945 heterarchy paper. Redraw the preference circuit as a directed
graph, prove the scalar-potential obstruction, and then return to McCulloch's
topological language of dromes, diadromes, and surface embedding.
\cite{McCulloch1945} The modern theorem is a foothold, not a replacement for the
paper's stronger biological speculation.

Read ``How We Know Universals'' with two columns of notes: averaging/scansion
and feedback/standardization.\cite{PittsMcCulloch1947} For every proposed
invariance, identify the transformation group, the feature that should remain
selective, and the physical cost of traversing or correcting the orbit. Follow
with ``Why the Mind Is in the Head'' to see reverberation, information,
prediction, and sensorimotor feedback in McCulloch's own synthesis.
Remember the chronology: the lecture was delivered at the September 1948 Hixon
Symposium and published in the 1951 proceedings.\cite{McCulloch1951}

Finally read the 1959 frog paper from behavior to anatomy, not only the four
findings paragraphs.\cite{Lettvin1959} For each fiber class, list the positive
stimuli, null stimuli, illumination manipulations, temporal persistence, field
size, conduction class, and tectal depth. Compare Hartline, Barlow, and Kuffler
first so the paper's advance is not mistaken for the invention of receptive
fields.\cite{Hartline1938,Hartline1940,Barlow1953,Kuffler1953} Then read one
modern retinal review to learn which conclusions endured and which taxonomy
expanded.\cite{GollischMeister2010,Donner2020}

For historical interpretation, use Piccinini for the computational claim,
Abraham for the intellectual and institutional context, and Arbib for the
experimental-epistemology arc. Gefter's \textit{Nautilus} profile is the
recommended biographical entry point for Pitts; read its vivid narrative beside
the more archival histories.
\cite{Piccinini2004,Abraham2002,Abraham2016,Arbib2000,Gefter2015}
Secondary history should discipline priority claims; it should not substitute
for working through the primary equations and experiments.

\section{From fixed logic to learnable neural systems}
\label{sec:bridge}

The unresolved task is to explain how a physical neural system learns the
right effective variables and invariances while remaining dynamically closed
under its own behavior. Rosenblatt's perceptron adds adaptive couplings but
largely takes its features as given. Amari asks when microscopic networks admit
macroscopic closure and which geometry makes learning intrinsic. Hopfield shows
how symmetry can turn recurrent dynamics into an energy landscape, at the cost
of excluding much nonequilibrium organization. These are not departures from
the logical calculus so much as attempts to make its circuit-level synthesis
learnable, statistical, and self-organizing. A theoretical NeuroAI must join
learning, recurrence, invariance, identifiability, and closed-loop control at a
scale that is both physically realizable and sufficient for the computation
being claimed.

\section{Conclusion}

\paragraph{Established result.}
McCulloch's legacy is smaller than the slogan that he ``invented neural
networks'' and larger than a cartoon threshold unit. The 1943 calculus of
McCulloch and Pitts made neural activity a causal formal system: threshold and inhibitory
veto realized propositions, delays realized temporal reference, and recurrent
circles realized internal state. A single threshold element is limited to
linearly separable functions, while finite networks can synthesize arbitrary
finite Boolean relations. The 1945 heterarchy argument connected directed
cycles to the failure of a global scalar ordering; the 1947 universals paper
made transformation invariance a circuit problem; and the 1959 frog experiment
identified four parallel, retinotopically registered operations through
single-fiber physiology and controlled stimulus variation.

\paragraph{Surviving principle.}
The durable inheritance is a standard of explanation. Specify the physical
degrees of freedom and their timescale; write the causal update law; state the
symmetry, equivalence class, or nuisance variation that defines the operation;
determine what feedback adds to the state space; and choose an intervention or
stimulus ensemble that makes the proposed computation identifiable. The
coarse-grained variables must also close: convenience does not make a binary
point-neuron state sufficient. This standard connects the formal synthesis of
the early papers to the experimental epistemology of the frog study without
claiming that their models or methods are identical.

\paragraph{Limitation.}
The 1943 model contains no general data-driven learning rule, no explicit
theory of noise tolerance, and no unbounded memory resource. A fixed recurrent
network with $N$ binary units is a finite-state machine, not an unbounded Turing
machine. Its absolute inhibition, synchronous unit delay, and binary state can
fail as effective variables when dendritic voltage, adaptation, stochasticity,
or heterogeneous conduction changes the future at fixed $\bm{x}(t)$. Likewise,
orbit averaging can destroy selectivity, feedback canonicalization can be
nonunique, and the supplied retinal channels are qualitative operation-level
models rather than fits to the original spike trains. Those limits define the
domain of the theory; they do not erase its causal and organizational content.

\section*{Data $\&$ Code Availability}

Code and figure-generation script are available at: 
\url{https://github.com/neurovium/NeuroAI}.

\section*{References}

\bibliography{mcculloch_book_chapter_references}
\end{document}